\documentstyle[preprint,aps,epsfig]{revtex}
\def\x{{\bf x}}
\def\y{{\bf y}}
\def\r{{\bf r}}
\def\a{{\alpha}}
\begin{document}
\draft

\title{The Complex Kohn Variational Method applied to N-d scattering} 

\author{ A.~Kievsky}
\address{ Istituto Nazionale di Fisica Nucleare, Piazza Torricelli 2,
          56100 Pisa, Italy }

\date{\today}

\maketitle

\abstract{ The three-nucleon ground state and the N--d scattering states 
           are obtained using variational principles.
           The wave function of the system
           is decomposed into angular--spin--isospin channels and
           the corresponding two dimensional spatial amplitudes are expanded in
           a correlated polynomial basis.
           For the scattering states, the complex form of the
           Kohn variational principle is used to determine the $S$--matrix.
           Special attention is given to the convergence pattern of the 
           phase--shift and mixing parameters. 
           The calculations have been performed using realistic 
           local NN potentials and three--nucleon forces. Important
           features of the method are anomaly--free solutions and
           the low dimensionality of the 
           matrices involved allowing for the inclusion of a large number
           of states. Very precise and stable numerical results have been
           obtained.} 

\bigskip

\noindent PACS numbers 03.65.Nk, 21.45.+v, 25.10.+s 

\bigskip

\noindent key words: three-nucleon system, N-d scattering, complex Kohn 
  variational principle, correlated hyperspherical harmonic basis, phase 
  shift and mixing parameters, $S$--matrix.

\bigskip

\noindent Alejandro Kievsky, Dipartimento di Fisica,
 Piazza Torricelli 2, 56100 Pisa - ITALY \\
tel: +39-50-911288, Fax: +39-50-911313, e-mail: kievsky@pisa.infn.it

\newpage

\section{Introduction}
 
Variational principles can be used to obtain the scattering states of
few--body systems. Among them, the  Kohn variational principle has recently
received renewed attention. In its usual form, the Kohn--Hulth\'{e}n method
is applied to calculate the reactance $K$--matrix or its inverse $K^{-1}$. In
ref~\cite{Schwartz} it was pointed out that the presence of singularities 
in the method was related to a zero eigenvalue in the spectrum of the operator 
$H-E$. When the operator $H-E$ is expanded in a real set of basis functions, the
associated matrix presents a number of eigenvalues close to zero and this number
increases with the dimension of the basis.
In order to avoid
these anomalous singularities the Kohn variational principle was
implemented for the $S$--matrix or $T$--matrix using proper boundary 
conditions. In this case the related quantities are complex matrices and so 
the singularities can be shifted to complex energies.
Occasionally, singularities can also occur in the complex Kohn method, 
as was revealed in ref.~\cite{Lucc}. But they can be eliminated with a
judicious choice of the non--linear variational parameters.
Moreover, the numerical check of the unitarity of $S$ can be used
to detect an anomalous behavior.
Applications of the complex Kohn variational method have been made
in atomic and molecular physics
for electron--atom and electron--molecule collisions~\cite{atomic} and
in nuclear physics in nucleon--nucleon (NN) scattering~\cite{Tomio}.
The principal benefits of the method are anomaly--free solutions and 
the simplicity in the computation of the required matrix elements.

An important problem in variational calculations is the study of the 
convergence pattern. When the wave function of the system
is expanded on a certain basis, in general a very large number of 
elements are needed to get fully converging results. 
This is particularly so in nuclear physics since the potential
contains a large repulsion at short distances and is strongly state--dependent.
Consequently, many states with different spin, isospin and angular momentum 
quantum numbers must be included in a single calculation.
In ref.~\cite{KVR94} the usual Kohn variational principle for the 
$K$--matrix was used to describe N--d scattering below the deuteron
breakup. The $S$--matrix was then obtained from the relation
\begin{equation}
S=(1+iK)(1-iK)^{-1}.
\end{equation}
However, as was pointed out in ref.~\cite{atomic}, it is not equivalent to 
applying the complex form of the principle directly to the $S$--matrix since 
the above relation holds only for the exact matrices.

In the present paper the complex form of the Kohn variational principle 
is applied to calculating N--d scattering processes. 
Using the technique of ref.~\cite{KVR94} the three-nucleon wave function
is decomposed into channels labeled by
spin, isospin and angular momenta quantum numbers. The corresponding 
two--dimensional spatial amplitudes are expanded in terms of the Pair
Correlated Hyperspherical Harmonic (PHH) basis~\cite{KVR93}.
The remaining unknown hyperradial functions are taken as a product of a
linear combination of Laguerre polynomials and an exponential tail. 

Realistic interactions are considered for investigating the flexibility of
the basis used to describe the structure the system. 
The Argone AV14~\cite{AV14} nucleon--nucleon (NN) potential is
used together with three-nucleon interaction (TNI) terms. 
The point Coulomb potential is included in the two proton
interaction.  In section 2 the three-nucleon bound states are calculated
as a first application of the polynomial basis. Useful information about
the number of basis elements needed to reach convergence and the
corresponding numerical accuracy is obtained. 
Section 3 is devoted to the description of N--d scattering 
using the complex Kohn variational principle. 
Scattering lengths and phase shift and mixing parameters have been 
calculated. In the N--d scattering, 
states with total angular momentum and parity
$J^\pi=1/2^+,3/2^+,1/2^-,3/2^-,5/2^-$ have been considered at the nucleon
incident energy $E_N=3.0$ MeV (just below the deuteron breakup).
Special attention is given
to the convergence of the elements of the $S$--matrix. 
Variations with respect to the nonlinear
parameters and the fulfillment of the unitarity condition $SS^\dagger=I$ 
are discussed.
The conclusions and perspectives of the method are given in the last section.

\section{Bound state calculations}

Following ref.~\cite{KVR93} the three--nucleon bound state wave function
is written as a sum of three Faddeev--like amplitudes
\begin{equation}
 \Psi = \psi(\x_i,\y_i)+\psi(\x_j,\y_j)+\psi(\x_k,\y_k)\ ,
\label{eq:Psi}
\end{equation}
where $\x_i,\y_i$ are the internal Jacobi coordinates which 
are defined in terms of the particle coordinates as
\begin{equation}
 \x_i=\r_j-\r_k\, ,\quad \y_i={1\over {\sqrt3}}(\r_j+\r_k-2\r_i)\ .
\end{equation}
Each $i$--amplitude has total angular momentum $JJ_z$ and total
isospin $TT_z$. Using $LS$ coupling it can be decomposed into channels
\begin{eqnarray}
     \psi(\x_i,\y_i) &=& \sum_\alpha^{N_c} \phi_\alpha(x_i,y_i) 
     {\cal Y}_\alpha (jk,i)  \\
     {\cal Y}_\alpha (jk,i) &=&
     \Bigl\{\bigl[ Y_{\ell_\alpha}(\hat x_i)  Y_{L_\alpha}(\hat y_i) 
     \bigr]_{\Lambda_\alpha} \bigl [ s_\alpha^{jk} s_\alpha^i \bigr ]
     _{S_\alpha}
      \Bigr \}_{J J_z} \; \bigl [ t_\alpha^{jk} t_\alpha^i \bigr ]_{T T_z},
\end{eqnarray}
where $x_i,y_i$ are the moduli of the Jacobi coordinates. 
Each $\alpha$--channel is labeled by the angular momenta $\ell_\alpha,
L_\alpha$ coupled to $\Lambda_\alpha$ and by the spin (isospin) 
$s_\alpha^{jk}$ ($t_\alpha^{jk}$) of the pair $j,k$ coupled to the spin
(isospin) of the third particle $s_\alpha^i$ ($t_\alpha^i$) to give 
$S_\alpha$ ($T$). $N_c$ is the number of channels taken into account in the
construction of the wave function and can be increased until convergence
is reached. The antisymmetrization of the state requires that
$\ell_\alpha+s_\alpha^{jk}+t_\alpha^{jk}$ be odd, while the parity of
the state is given by $\ell_\alpha+L_\alpha$. 

The two--dimensional spatial amplitude in each channel is expanded in
terms of the PHH basis:
\begin{equation}
     \phi_\alpha(x_i,y_i) = \rho^{\ell_\alpha+L_\alpha} f_\alpha (x_i)
     \left[ \sum_K u^\alpha_K(\rho) {}^{(2)}P^{\ell_\alpha,L_\alpha}_K(\phi_i)
     \right] \ ,
\label{eq:PHH}
\end{equation}
where the hyperspherical polynomials are 
\begin{equation}
    {}^{(2)}P^{\ell_\alpha,L_\alpha}_K(\phi_i)=N_n^{\ell_\alpha,L_\alpha}
    (\sin\phi_i)^{L_\alpha}(\cos\phi_i)^{\ell_\alpha}
    P_n^{L_\alpha+1/2,\ell_\alpha+1/2}(\cos2\phi_i) \ .
\end{equation}
Here,
$N_n^{\ell_\alpha,L_\alpha}$ is a normalization factor, $P_n^{\alpha,\beta}$
is a Jacobi polynomial and $K=\ell_\alpha+L_\alpha+2n$ is the grand orbital
quantum number which runs from its minimum value $K_0=\ell_\alpha+L_\alpha$
to its maximum selected value $K_\alpha$. Therefore, the number of
hyperradial functions per channel is $M_\alpha=(K_\alpha-K_0)/2+1$.
The inclusion of the pair correlation function $f_\alpha(x_i)$ in 
the expansion of eq.(\ref{eq:PHH}), as has been discussed in ref.~\cite{KVR93},
accelerates the convergence taking into account
the correlations introduced by the strong repulsion of the NN potential.
It is obtained solving a two--body Schr\"odinger--like equation.

In the present work the hyperradial functions $u^\alpha_K(\rho)$ are taken
as a product of a linear combination of Laguerre polynomials 
and an exponential tail:
\begin{equation}
 u^\alpha_K(\rho)=\sum_m A^\alpha_{K,m}L^{(5)}_m(z)\exp(-z)\ ,
\label{eq:hfun}
\end{equation}
where $z=\gamma\rho$ and $\gamma$ is a nonlinear variational parameter.
Let $|\alpha,K,m>$ be a totally antisymmetric element of the expansion 
basis, where $\alpha$ denotes the channel including
the angular-spin-isospin dependence and the correlation function,
and $K,m$ are the indices of the 
hyperspherical and Laguerre polynomials, respectively. In terms of the basis
elements the wave function (\ref{eq:Psi}) can be writen as
\begin{equation}
 \Psi=\sum_{\alpha,K,m}A^\alpha_{K,m}|\alpha,K,m>.
\end{equation}
The problem is to determine the linear coefficients $A^\alpha_{K,m}$.
To this aim, the Hamiltonian considered here includes the AV14 NN 
potential and the three--body Tucson--Melbourne (TM) force~\cite{TM}.
The cutoff parameter of the TM potential is chosen to be $\Lambda=5.13\mu$
($\mu$ is the pion mass)
in order to reproduce the $^3$H binding energy.
To evaluate Coulomb (C) effects the AV14+C potential model is also analyzed.
The choice of the interaction model allows for rigorous comparisons with other
techniques and it reproduces reasonably well the structure of the 
three--nucleon system.

The wave function and the
energy of the system are obtained by solving the following 
generalized eigenvalue problem
\begin{equation}
\sum_{\alpha',K',m'}A^{\alpha'}_{K',m'}
    <\alpha,K,m|H-E|\alpha',K',m'>=0 \ .
\label{eq:matrix}
\end{equation}
In the latter equation the dimension of the involved matrices
is related to three
indices: the number of $\alpha$--channels $N_c$, the number of hyperspherical
polynomials for each channel $M_\alpha$ and the number of Laguerre polynomials
included in the expansion of the hyperradial functions (\ref{eq:hfun}). 
To analyze the convergence properties of the expansion, the indices $K,m$
have been increased until stable results are reached.
Beginning with those channels having minimum grand angular quantum number
$K_0=0$, new channels with higher $K_0$ values are successively included.
The ground state has total angular
momentum and parity $J^\pi=1/2^+$ and total isospin $T=1/2$. The number
of channels with $K_0\le 4$ is $N_c=18$ and with $K_0\le 6$ is $N_c=26$.
In refs.~\cite{KVR94,KVR93} it was pointed out that the
number of hyperspherical functions per channel needed when using the PHH
is small: typically
$6$ to $8$ functions in those channels with the deuteron quantum numbers
$(\ell_\alpha=0,2;s_\alpha^{ij}=1;t_\alpha^{ij}=0)$,
and $4$ to $6$ in the other ones are sufficient for a four digit accuracy.

The matrix elements of eq.(\ref{eq:matrix}) 
have been calculated numerically using a Gauss formula for integrals
in the variable $\mu_i={\hat x}_i \cdot {\hat y}_i$, 
a Chebyshev Lobatto formula for integrals
in the variable $\cos{2\phi_i}$ and a Lagrange formula for the
integrals in the hyperradius $\rho$~\cite{abramowitz}.

Special attention has been paid to the study of the convergence with the 
Laguerre polynomials and the nonlinear parameter $\gamma$.
The results for the three potential models are indicated in fig. 1. In each
panel three curves are given corresponding to the number of channels
$N_c=14,18,26$ (the ordering of the channels is given, for instance, in
ref.~\cite{Bench1}). 
The binding increases with the number of channels, but 
changing $N_c$ from 18 to 26 produces an increment of about $1$ keV. Higher
channels are strongly suppressed due to centrifugal barrier effects.
The $N_c=26$ results for the binding energy, kinetic energy and 
$S$--, $D$-- and $P$--wave occupation percentage are given in table 1.
High precision results from refs.~\cite{Bench1,Bench2} are also shown
for the sake of comparison. An extremely close agreement has been obtained
in all the calculated quantities.
In the final calculation we have $N_c=26$, $m=16$ and $\gamma=2.0$ fm$^{-1}$.
The number of basis elements was $1900$, which is the dimension 
of the matrices in the generalized eigenvalue problem.

The variation of the binding energy $B$ with the nonlinear parameter
$\gamma$ and with the number of Laguerre polynomials $m=4,8,12,16$ 
is presented in fig.2. It can be seen that the dependence on $\gamma$
is appreciable only for rather low $m$ values. On the other hand,
the convergence with $m$ is very fast and for $m \ge 12$
the results are already accurate at the $0.1\%$ level. The number
of basis elements is large enough to guarantee a weak dependence with the
nonlinear parameter.

The results obtained so far show the great flexibility of the basis for 
reproducing the structure of the system, even when the AV14+TM and AV14+Coulomb
potential models are used. 
There are no difficulties in applying the polynomial basis to
other realistic NN interactions such as the Argonne AV18 potential, 
Bonn potential, Nijmegen potential, etc. 
For those interactions expressed in momentum space
the matrix elements of eq.(\ref{eq:matrix}) 
can be calculated after performing a Fourier transformation the basis. 

\section{The complex Kohn Variational Principle}

Following ref.~\cite{KVR94}, the wave function corresponding to an N-d
scattering state is written as a sum of two terms
\begin{equation}
\Psi=\Psi_C +\Psi_A \ .
\end{equation}
The $\Psi_C$ term describes the system when the three--nucleons are close to
each other. For large interparticle separations and energies below the
deuteron breakup it goes to zero, whereas for higher energies it must
describe three outgoing particles~\cite{KVR97}. The second term $\Psi_A$
is the
solution of the Schr\"odinger equation in the asymptotic nuclear region, where
the two incident clusters are well separated. It can be written as a linear
combination of functions of the form
\begin{eqnarray}
   \Omega^\lambda_{LSJ}(\x_i,\y_i) & = \sum_{l_\a=0,2} w_{l_\a}(x_i)
       {\cal R}^\lambda_L (y_i) \times \nonumber \\
       &\left\{\left[ [Y_{l_\a}({\hat x}_i) s_\a^{jk}]_1 \otimes s^i \right]_S
       \otimes Y_L({\hat y}_i) \right\}_{JJ_z}
       [t_\a^{jk}t^i]_{TT_z}\ , \label{eq:omega}
\end{eqnarray}
where $w_{l_\a}(x_i)$ is the deuteron component in the state
$l_\a =0,2$ and $L$ is the relative angular momentum of the deuteron and the
incident nucleon. For $\lambda \equiv R$, ${\cal R}^\lambda_L (y_i)$
coincides with the regular solution $F_L(y_i)$ of the two--body N--d
Schr\"odinger equation in absence of nuclear interaction.
For $\lambda \equiv I $, it is set equal to ${\widetilde G}_L(y_i)$,
the product of the
irregular solution $G_L(y_i)$ and a regularizing factor which goes to
zero as $y_i^{L+1}$, when $y_i\rightarrow 0$, 
and approaches unity for large values of $y_i$:
\begin{equation}
{\widetilde G}_L(y)=(1-{\rm e}^{-\xi r_{Nd}})^{L+1}G_L(y),
\end{equation}
where $r_{Nd} =(2/ \sqrt3)y$ is the nucleon--deuteron separation. 
Hereafter the parameter $\xi$ is given the value $\xi=0.25$ 
fm$^{-1}$~\cite{KVR94}.
The asymptotic
wave function can also be  constructed in terms of the physical ingoing
and outgoing solutions by linearly combining the above functions.
As in ref.~\cite{Lucc} we can define a general asymptotic state 
\begin{equation}
\Omega^+_{LSJ}(\x_i,\y_i) =  \Omega^0_{LSJ}(\x_i,\y_i)+
 \sum_{L'S'}{}^J{\cal L}^{SS'}_{LL'}\Omega^1_{L'S'J}(\x_i,\y_i)  \ ,
\end{equation}
with the following asymptotic functions
\begin{eqnarray}
\Omega^0_{LSJ}(\x_i,\y_i) =& u_{00}\Omega^R_{LSJ}(\x_i,\y_i)+
                            u_{01}\Omega^I_{LSJ}(\x_i,\y_i) \ , \\
\Omega^1_{LSJ}(\x_i,\y_i) =& u_{10}\Omega^R_{LSJ}(\x_i,\y_i)+
                            u_{11}\Omega^I_{LSJ}(\x_i,\y_i)  \ ,
\end{eqnarray}
Four different choices of the matrix ${\cal L}$ can be obtained 
in correspondence to the following definitions for the matrix $u$.
\begin{eqnarray}
{\cal L} & =K     \quad{\rm for}\quad u= \left( \begin{array}{cc}
                                   1 & 0 \\ 0 & 1\end{array}\right) \ , \\
{\cal L} & =K^{-1}\quad{\rm for}\quad u= \left( \begin{array}{cc}
                                   0 & 1 \\ 1 & 0\end{array}\right) \ , \\
{\cal L} & =S     \quad{\rm for}\quad u= \left( \begin{array}{cc}
                                   i &-1 \\ i & 1\end{array}\right) \ , \\
{\cal L} & =-\pi T\quad{\rm for}\quad u= \left( \begin{array}{cc}
                                   1 & 0 \\ i & 1\end{array}\right)  \ .
\end{eqnarray}
Accordingly, the asymptotic states have been normalized in order to
satisfy the Wronskian
\begin{equation}
W(\Omega^0_{LSJ},\Omega^1_{LSJ})=u_{01}u_{10}-u_{00}u_{11}=-{\rm det}(u) \ .
\end{equation}

The three-nucleon scattering wave function for an incident 
state with relative angular momentum $L$, spin $S$ and total angular momentum
$J$ is
\begin{equation}
\Psi^+_{LSJ}=\sum_{i=1,3}\left[ \Psi_C(\x_i,\y_i)+\Omega^+_{LSJ}(\x_i,\y_i)
             \right] \ ,
\end{equation}
and its complex conjugate is $\Psi^-_{LSJ}$. A variational estimate for the
matrix ${\cal L}$ can be obtained from the generalized Kohn variational
principle
\begin{equation}
[{}^J{\cal L}^{SS'}_{LL'}]= {}^J{\cal L}^{SS'}_{LL'}-{2\over {\rm det}(u)}
\langle\Psi^+_{LSJ}|H-E|\Psi^+_{L'S'J}\rangle \ .
\label{eq:kohn}
\end{equation}
The trial parameters in the wave function $\Psi^+_{LSJ}$ are varied in order
to obtain a stationary value of the above functional.

In the present work the calculations have been restricted to energies
below the three-body breakup. Therefore, the $\Psi_C$ term goes to zero
when the hyperradius $\rho\rightarrow\infty$, and it has been expanded in terms
of the polynomial basis introduced in the previous section.
\begin{equation}
 \Psi_C=\sum_{\alpha,K,m}A^\alpha_{K,m}|\alpha,K,m>.
\end{equation}
With this choice of $\Psi_C$, the variation of the diagonal functionals
with respect to the linear parameters leads to the following linear system
\begin{equation}
\sum_{\alpha',K',m'}A^{\alpha'}_{K',m'}
    <\alpha,K,m|H-E|\alpha',K',m'>=D^\lambda_{LSJ}(\alpha,K,m) \ ,
\label{eq:linear}
\end{equation}
with the two different inhomogeneous $D$ terms corresponding to
$\lambda\equiv 0,1$
\begin{equation}
 D^\lambda_{LSJ}(\alpha,K,m)=\sum_{j}
    <\alpha,K,m|H-E|\Omega^\lambda_{LSJ}(\x_j,\y_j)> \ .
\end{equation}
The first order solution of the matrix ${\cal L}$ 
is obtained by solving the following algebraic equations
\begin{equation}
\sum_{L'',S''}{}^J{\cal L}^{SS''}_{LL''} X^{S'S''}_{L'L''}= Y^{SS'}_{LL'} \ ,
\end{equation}
with the coefficients $X$ and $Y$ defined as
\begin{eqnarray}
X^{SS'}_{LL'}=& \langle\Omega^1_{LSJ}+\Psi^1_{LSJ}|H-E|\Omega^1_{L'S'J}\rangle
 \ ,\\
Y^{SS'}_{LL'}=& \langle\Omega^0_{LSJ}+\Psi^0_{LSJ}|H-E|\Omega^0_{L'S'J}\rangle
 \ ,
\end{eqnarray}
where $\Psi^\lambda_{LSJ}$ is the solution of the set of eqs.(\ref{eq:linear})
with the corresponding inhomogeneous term. The second order estimate 
$[{}^J{\cal L}^{SS'}_{LL'}]$
is obtained from the first order solution using eq.(\ref{eq:kohn}).

The calculations has been performed applying eq.(\ref{eq:kohn}) 
to the $S$--matrix.
The unitarity of the $S$--matrix, $SS^\dagger=I$, has been checked to
verify the completeness of the basis and to detect eventual spurious
solutions. The reactance $K$--matrix, which is real and symmetric,
can be obtained from the relation 
\begin{equation}
    K=(u_{01}+u_{11}{\cal L})(u_{00}+u_{01}{\cal L})^{-1} \ .
\end{equation}

In an N--d process, the total energy of the system is $E=-B_d+E_0$
with $B_d$ the binding energy of the deuteron and $E_0$ 
the incident nucleon energy. As a first application of the method,
the zero energy process at $E_0=0$ is studied.
In this case only $S$--wave scattering has to be considered, with two
possible states, $J=1/2^+$ and $J=3/2^+$. The Hamiltonian of the system
includes two and three--nucleon forces. As for the bound state,
the AV14, the AV14+TM and AV14+C potential models are utilized. 
After fixing the number of channels $N_c$, eq.(\ref{eq:linear})  has been
solved with increasing values of $K,m$.  States up to $K_0\le 6$ have been
taken into account, which means $N_c=26$ for $J=1/2^+$ and $N_c=44$ for
$J=3/2^+$. The second order results for the doublet $^2a$ and quartet $^4a$ 
scattering lengths and the difference $|SS^\dagger -I|$ are given in table 2
for a few values of the number $m$ of Laguerre polynomials.
In this process the $S$--matrix is a scalar, therefore to check the unitarity 
condition the square of the modulus of a complex number must be calculated.
The difference between the first and second order results measures
the error in the construction of the wave function. To give an idea,
the relative difference is $0.1$ for $m=4$, but it is $10^{-4}$ for $m=20$.
The accuracy of the solution is also reflected in the unitarity condition 
which is satisfactorily fulfilled, 
indicating the absence of spurious solutions~\cite{Lucc}.

N--d scattering at zero energy has been studied
using similar potential models in ref.~\cite{sctl},
solving the corresponding Faddeev equations in configuration
space, and in ref.~\cite{KVR94} applying the Kohn variational principle to
the $K$--matrix. The corresponding results are given in
the table for the sake of comparison. The differences can be attributed to
working in a different Hilbert space. 
For example, in ref.~\cite{sctl} a truncation
of the potential retaining those NN partial waves with $j\le4$ has been
considered, whereas in ref.~\cite{KVR94} a smaller number of channels in the
expansion of the wave function has been utilized.

The calculations reported the table 2 have been done with a value of
the nonlinear parameter $\gamma=1.5$ fm$^{-1}$. The results for $^2a$ 
when $\gamma$ is varied
around this value are shown in the first three panels of fig.3,
for the three potential models under consideration. A similar behavior is
obtained for $^4a$, with the conclusion that the results have only a
slight dependence on $\gamma$ when enough polynomials are included.
It is interesting to note that in this scattering calculation
the convergence is faster around the value of $\gamma=1.5$ fm$^{-1}$,
whereas in the bound state case this occurs around $\gamma=2.0$
fm$^{-1}$. The lower value of $\gamma$ is motivated by the fact that the
expansion basis must construct the scattering 
wave function (\ref{eq:Psi}) also at intermediate
particle separations where the term $\Psi_A$ is not yet the correct solution.
Finally, in the last panel of fig.3 the convergence with the number
of channels is displayed. Changing the number of channels from $N_c=18$ to
$N_c=26$ produces a variation of the scattering length less than
$10^{-3}$ fm.

For positive incident energies, $S$--wave and $P$--wave scattering
have been studied. Accordingly, five different states can be constructed
with $J=1/2^+,3/2^+,1/2^-,3/2^-,5/2^-$. When $J=1/2$ the corresponding
$S$--matrix is a $2\times 2$ matrix, in all the other cases it is a 
$3\times 3$ matrix. 
For each $J^\pi$ state all channels with $K_0\le 6$ have been included. 
It was checked that higher channels give negligible contributions.
The case considered here corresponds to a nucleon incident center of mass
energy of $E_0=2.0$ MeV ($E_{lab}=3.0$ MeV), slightly below
the deuteron breakup. 
The results of main interest are presented in figs.4, 5. In fig.4
the two $S$--wave phase shifts and the five $P$--wave phase shifts
are shown for different numbers of Laguerre polynomials $m$, whereas
the corresponding mixing parameters are presented in fig.5.
The solid curve corresponds to the 
AV14 potential model, the dashed line to the AV14+C and
the dotted line to the AV14+TM potential models, respectively. 
Stable results are obtained for $m > 8$, especially 
for the mixing parameters. An increase of $m$ from $16$ to $20$ gives
no appreciable contribution.  More interesting is the constant 
difference between the AV14 and the other two potential models thereby
providing a fast method for evaluating Coulomb and TNI effects.

For a more detailed analysis of the convergence properties of the correlated
basis in conjunction with the complex Kohn principle, the $S$--matrix 
elements $S^{SS'}_{LL'}$ are given for the $J=3/2^-$ state in table 3.
This case illustrates the main aspects of the convergence
with similar patterns in the other four cases. 
With the condition $K_0\le 6$ the number of channels
allowed are $N_c=36$, which suffices for a four digit accuracy. In the table,
the second order solutions for the three phase shifts and the three mixing 
parameters of the state, are given in columns $2$--$7$.
The absolute value of the diagonal elements of the matrix $SS^\dagger-I$ 
are also shown in the last three columns.
The $S$--matrix is ordered following
the convention of Seyler~\cite{Seyler}, therefore the first diagonal element
corresponds to the $^4F$ state, the second one to the $^2P$ state and
the third one to the $^4P$ state. For all the six parameters the
convergence is quite satisfactory and the unitarity condition is fulfilled.
The complicated structure of the TNI is reflected in a slightly
worse, although satisfactory fulfillment of this condition.
The behaviour of the phase shift and mixing parameters with the
nonlinear parameter $\gamma$ is similar to that found for the bound 
state and for the scattering lengths. Again a faster convergence is obtained
around the value $\gamma=1.5$ fm$^{-1}$ as in the preceding case.
The above analysis can be repeated for states with different 
$J$ and parity values, yielding in all cases quite accurate results.

\section{Conclusions}

In the present paper a detailed numerical analysis of the three--nucleon bound
state and of the N--d scattering have been presented. For the N--d process
the complex Kohn variational principle has been utilized, limiting the study 
to energies below the deuteron breakup. 
A correlated polynomial basis
with exponential tail has been used to expand the internal part
of the three--nucleon wave function. The Hamiltonian of the system
includes realistic two-- and three--nucleon interaction terms and the point
Coulomb potential. Explicitly, the Argonne AV14 potential and the
Tucson--Melbourne $\pi$--$\pi$ three--nucleon force have been considered.
Two different motivations have been pursued. First of all, the variational
method investigated allows for an accurate description of the scattering
states. Singularity--free solutions have been obtained and the unitarity
condition for the $S$--matrix is satisfied. The second motivation
was the study of the convergence pattern related to the use 
of the correlated polynomial basis. This basis
presents a fast rate of convergence and a good stability with the nonlinear
parameter $\gamma$ in the exponential tail. Binding energies, 
scattering lengths and phase--shift and mixing parameters have been
calculated with an error of less than $0.1\%$. The number of basis elements
is reasonably small and the related dimension of the matrices is kept low.

The analysis of the convergence was studied paying attention 
on three indices. 
For a given $J^\pi$ state, all channels corresponding to
$K_0=\ell_\alpha+L_\alpha \le 6$ have been taken into account. 
At the energies considered here, channels with higher
values of $K_0$ are strongly suppressed due to centrifugal barrier effects.
The number of correlated hyperspherical functions for each channel has been
increased until a complete stability in the calculated quantities was obtained.
In this analysis the bound state calculations, presented
in section 2, and previous experience from ref.~\cite{KVR94}
was utilized.  The conclusion is that 8 hyperspherical functions 
in those channels with the deuteron quantum numbers
and 4 to 6 in the others suffice for the desired accuracy. Finally,
the convergence with respect to
the number of Laguerre polynomials was carefully studied.
Increasing $m$ from $16$ to $20$ does not produce any numerical improvement
and the convergence is already achieved with $m=12$.
When the maximum number of basis elements is used,
the relative difference between
the first order and second order results for the $S$--matrix is $10^{-4}$
and the unitarity condition is fulfilled up to $10^{-9}$ (without TNI) and
$10^{-6}$ (with TNI). The difference reflects the complicated structure of the
system when the TNI is present.

With the method presented here, it will be possible to give accurate
theoretical predictions for different observables such as the
N-d differential cross section, vector and tensor analyzing powers, etc.
In particular, it is possible to use the present method for 
evaluating TNI effects in polarization observables where important
differences have been observed between theoretical calculations and
the available experimental data~\cite{KRTV96}. 
In conclusion, the extension of the method to treat the breakup 
channel is of great interest.  The complex Kohn variational principle
is well suited for including outgoing boundary conditions in the
internal part of the wave function to describe the three outgoing nucleons.
A first application of the method including the breakup channel is
given in ref.~\cite{KVR97} and is at present subject of intense investigations.

\section{Acknowledgements}

The author wish to thank S. Rosati and M. Viviani for many 
fruitful discussions, and L. Lovitch for a careful reading of the manuscript.

\newpage
\noindent{\bf Table captions}

\bigskip

\noindent{\bf Table 1.} Binding energy and kinetic energy (in MeV),
         and $S$--, $D$-- and $P$--wave
         occupation probabilities for the three potential models under
         consideration. 

\bigskip

\noindent{\bf Table 2.} The doublet and quartet scattering lengths (in fm) 
         for different values of the number $m$ of Laguerre polynomials and 
         for the three potential models considered. The
         unitarity condition is checked in terms of the absolute difference
         $|SS^\dagger-I|$.

\bigskip

\noindent{\bf Table 3.} Phase shift and mixing parameters corresponding 
to the state $J=3/2^-$ as functions of the number $m$ of Laguerre
         polynomials. In the last three columns the unitarity of the 
         $S$--matrix is checked reporting the diagonals elements of the
         matrix difference $|SS^\dagger-I|$.

\newpage

\noindent{\bf Figure captions}

\bigskip

\noindent{\bf Figure 1.} 
                The convergence of the binding energy (in MEV)
                is shown as a function of the number $m$ 
                of Laguerre polynomials. The solid line
                corresponds to the number of channels $N_c=26$ in the
 expansion of the wave function; the dot--dashed line to $N_c=18$ and
                the dashed line to $N_c=14$. 

\bigskip

\noindent{\bf Figure 2.} 
                The calculated value of the binding energy (in MeV)
                as a function of the nonlinear
                parameter $\gamma$. The different curves in each panel 
                correspond to increasing the number $m$ of Laguerre polynomials.
           The long dotted line corresponds to $m=4$, the short dotted line to
                $m=8$, the dashed line to $m=12$ and the solid line to $m=16$.

\bigskip

{\bf Figure 3.} The doublet scattering length (in fm) as a function of the
                nonlinear parameter $\gamma$ (first three panels) and
                for increasing number of channels (fourth panel). 
                In panels $1-3$ the curves are given for increasing number 
                $m$ of Laguerre polynomials, as in fig. 2. In panel 4 the
                convergence with respect to the number of channels 
                is shown for the three potential models considered.

\bigskip

{\bf Figure 4.} Convergence of the phase shift parameters ($J=3/2^-$)
                with respect to the number $m$ of Laguerre polynomials. 
                For the sake of clarity the AV14+C is given separately
                in panels 4 and 6.

\bigskip

{\bf Figure 5.} Convergence of the mixing parameters ($J=3/2^-$)
                with respect to the number $m$ of Laguerre polynomials.

\newpage

\begin{table}
\label{tab:binding}
\begin{tabular}{c|ccccc}
 potential  & B(MeV) & T(MeV) & $P_S$(\%) & $P_D$(\%) & $P_P$(\%)   \\ 
\hline
   AV14     &  7.684 & 45.678 &  90.957   &  8.967    & 0.076      \\
   ref.~\cite{Bench1}
            &  7.684 & 45.677 &  90.956   &  8.968    & 0.076      \\
\hline
   AV14+C   &  7.033 & 44.813 &  90.994   &  8.931    & 0.075      \\
   ref.~\cite{Bench1}
            &  7.033 & 44.812 &  90.993   &  8.932    & 0.075      \\
\hline
   AV14+TM  &  8.486 & 49.337 &  90.577   &  9.262    & 0.162      \\
   ref.~\cite{Bench2}
            &  8.486 & 49.340 &           &           &            \\
\hline
\end{tabular}
\caption{}
\end{table}

\begin{table}
\newcommand{\SS}{|SS^\dagger-I|}
\label{tab:sctl}
\begin{tabular}{c|cccc}
\hline
            &\multicolumn{4}{c}{AV14} \\
\hline
 $m$        &$^2a$ (fm)& $\SS$  & $^4a$ (fm) & $\SS$  \\
\hline
   4        &  1.426 & 10$^{-3}$&  6.399  & 10$^{-4}$ \\
   8        &  1.203 & 10$^{-4}$&  6.381  & 10$^{-6}$ \\
   12       &  1.190 & 10$^{-6}$&  6.379  & 10$^{-7}$ \\
   16       &  1.189 & 10$^{-8}$&  6.379  & 10$^{-8}$ \\
   20       &  1.189 & 10$^{-9}$&  6.379  & 10$^{-9}$ \\
\hline
 ref.~\cite{KVR94}    &  1.196 &          &  6.380  &           \\
 ref.~\cite{sctl}    &  1.204 &          &  6.380  &           \\
\hline
            &\multicolumn{4}{c}{AV14+TM} \\
\hline
 $m$        &$^2a$ (fm)& $\SS$  & $^4a$ (fm) & $\SS$  \\
\hline
   4        &   0.8541 & 10$^{-2}$&  6.391  & 10$^{-2}$ \\
   8        &   0.6016 & 10$^{-4}$&  6.373  & 10$^{-4}$ \\
   12       &   0.5869 & 10$^{-7}$&  6.371  & 10$^{-7}$ \\
   16       &   0.5858 & 10$^{-6}$&  6.371  & 10$^{-6}$ \\
   20       &   0.5857 & 10$^{-6}$&  6.371  & 10$^{-6}$ \\
\hline
            &\multicolumn{4}{c}{AV14+C} \\
\hline
 $m$        &$^2a$ (fm)& $\SS$  & $^4a$ (fm) & $\SS$  \\
\hline
   4        &  1.369   & 10$^{-2}$&  14.002 & 10$^{-2}$ \\
   8        &  0.9656  & 10$^{-4}$&  13.790 & 10$^{-4}$ \\
   12       &  0.9431  & 10$^{-5}$&  13.774 & 10$^{-6}$ \\
   16       &  0.9414  & 10$^{-8}$&  13.773 & 10$^{-8}$ \\
   20       &  0.9411  & 10$^{-9}$&  13.773 & 10$^{-9}$ \\
\hline
 ref.~\cite{KVR94}     &  0.954   &          &  13.779 &           \\
 ref.~\cite{sctl}    &  0.965   &          &  13.764 &           \\
\hline
\end{tabular}
\caption{}
\end{table}

\begin{table}
\label{tab:J3/2}
\begin{tabular}{c|ccccccccc}
\hline
       & \multicolumn{9}{c}{AV14} \\
\hline
 $m$   &$^4F_{3/2-}$&$^2P_{3/2-}$&$^4P_{3/2-}$&$\eta_{3/2-}$&
        $\epsilon_{3/2-}$&$\zeta_{3/2-}$&
        \multicolumn{3}{c}{$|SS^\dagger-I|_{jj}$} \\
\hline
  4&0.9403&-7.488&26.19&-3.656&-2.712&-0.2715 &$10^{-4}$&$10^{-3}$&$10^{-2}$ \\ 
  8&0.9425&-7.216&26.38&-3.794&-2.755&-0.2592 &$10^{-6}$&$10^{-7}$&$10^{-5}$ \\ 
 12&0.9427&-7.198&26.39&-3.804&-2.760&-0.2580 &$10^{-7}$&$10^{-7}$&$10^{-6}$ \\ 
 16&0.9428&-7.197&26.40&-3.804&-2.761&-0.2578 &$10^{-7}$&$10^{-7}$&$10^{-8}$ \\ 
 20&0.9428&-7.196&26.40&-3.804&-2.761&-0.2578 &$10^{-9}$&$10^{-9}$&$10^{-8}$ \\ 
\hline
       & \multicolumn{9}{c}{AV14+TM} \\
\hline
 $m$   &$^4F_{3/2-}$&$^2P_{3/2-}$&$^4P_{3/2-}$&$\eta_{3/2-}$&
        $\epsilon_{3/2-}$&$\zeta_{3/2-}$&
        \multicolumn{3}{c}{$|SS^\dagger-I|_{jj}$} \\
\hline
  4&0.9408&-7.456&26.47&-3.681&-2.816&-0.2588 &$10^{-4}$&$10^{-3}$&$10^{-2}$ \\ 
  8&0.9430&-7.183&26.68&-3.819&-2.859&-0.2465 &$10^{-6}$&$10^{-6}$&$10^{-4}$ \\ 
 12&0.9432&-7.166&26.69&-3.827&-2.865&-0.2448 &$10^{-7}$&$10^{-6}$&$10^{-6}$ \\ 
 16&0.9433&-7.163&26.69&-3.829&-2.866&-0.2448 &$10^{-7}$&$10^{-6}$&$10^{-6}$ \\ 
 20&0.9433&-7.163&26.69&-3.830&-2.866&-0.2448 &$10^{-8}$&$10^{-6}$&$10^{-6}$ \\ 
\hline
       & \multicolumn{9}{c}{AV14+C} \\
\hline
 $m$   &$^4F_{3/2-}$&$^2P_{3/2-}$&$^4P_{3/2-}$&$\eta_{3/2-}$&
        $\epsilon_{3/2-}$&$\zeta_{3/2-}$&
        \multicolumn{3}{c}{$|SS^\dagger-I|_{jj}$} \\
\hline
  4&0.8705&-7.353&24.45&-3.273&-2.295&-0.3301 &$10^{-4}$&$10^{-3}$&$10^{-2}$ \\ 
  8&0.8718&-7.175&24.61&-3.356&-2.323&-0.3350 &$10^{-6}$&$10^{-8}$&$10^{-5}$ \\ 
 12&0.8720&-7.167&24.62&-3.360&-2.327&-0.3219 &$10^{-6}$&$10^{-7}$&$10^{-6}$ \\ 
 16&0.8721&-7.166&24.63&-3.362&-2.327&-0.3219 &$10^{-6}$&$10^{-7}$&$10^{-8}$ \\ 
 20&0.8721&-7.166&24.63&-3.362&-2.327&-0.3219 &$10^{-8}$&$10^{-8}$&$10^{-8}$ \\ 
\hline
\end{tabular}
\caption{}
\end{table}

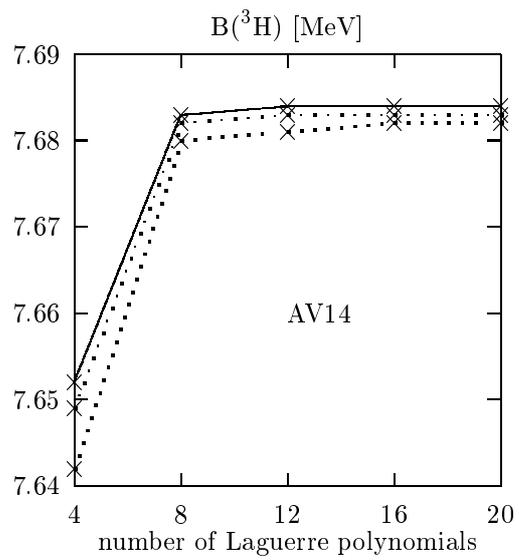
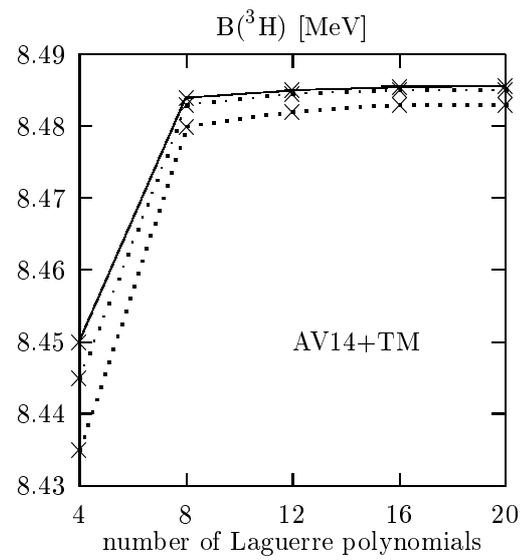
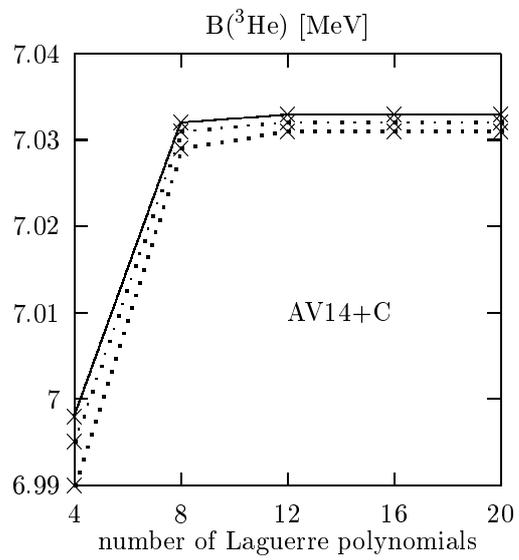
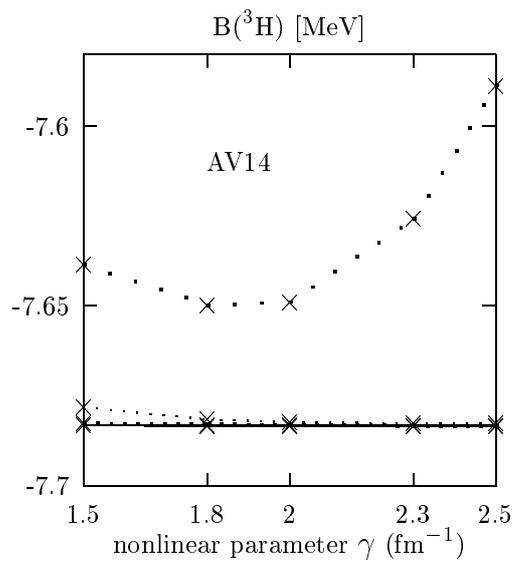
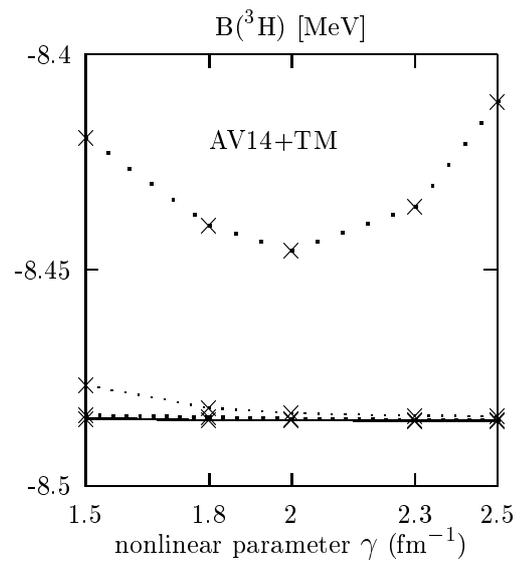
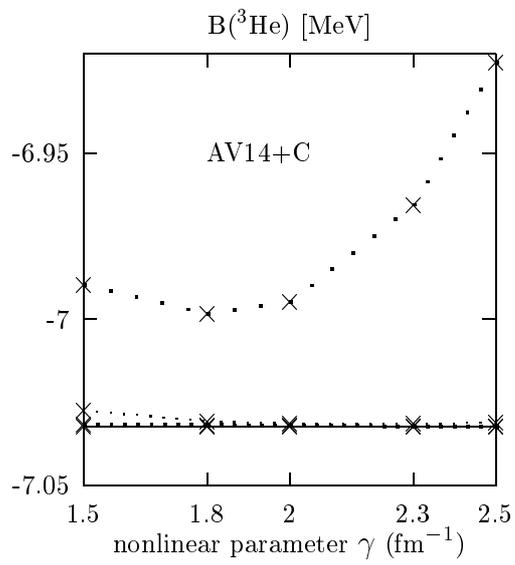
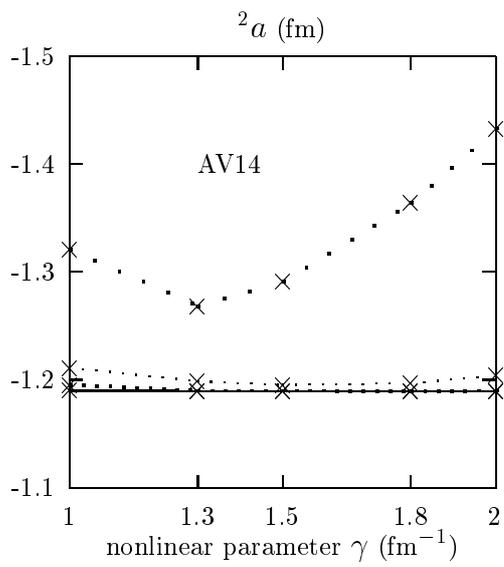
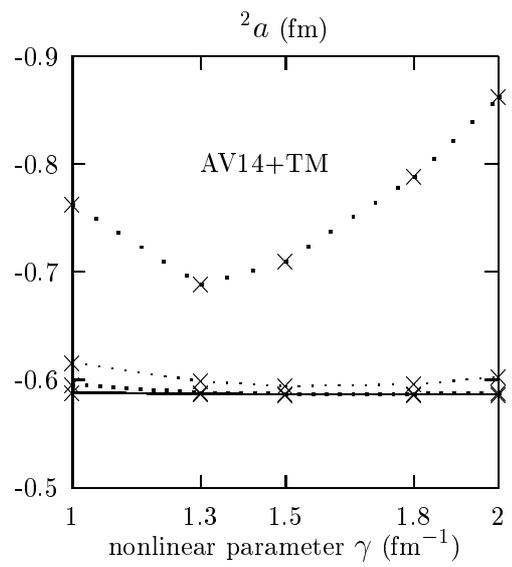
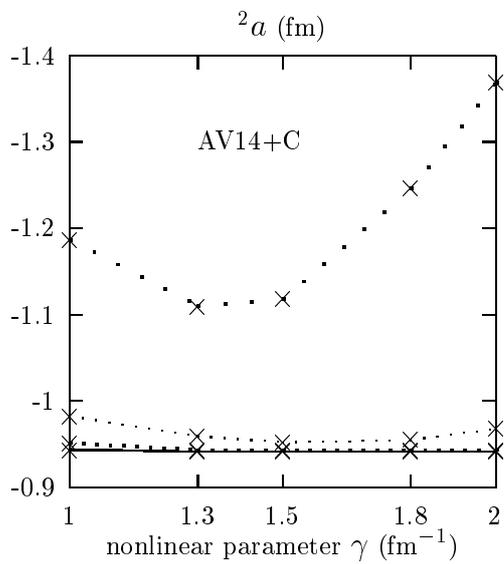
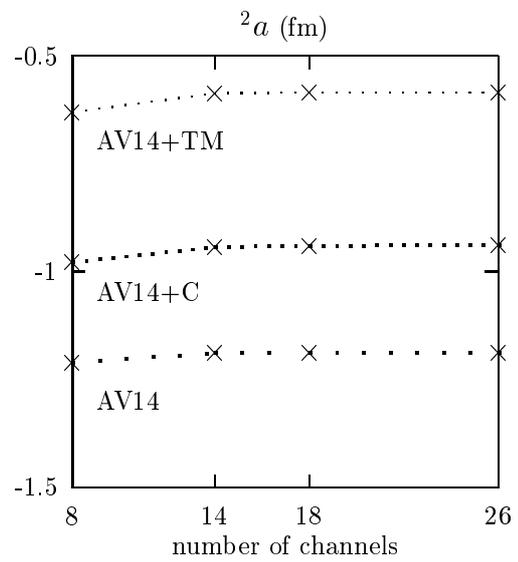
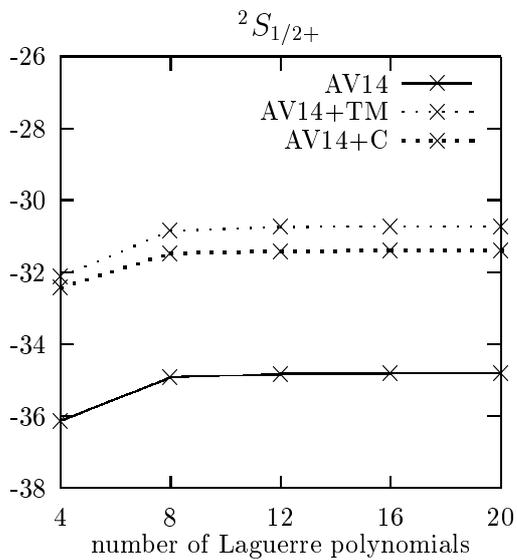
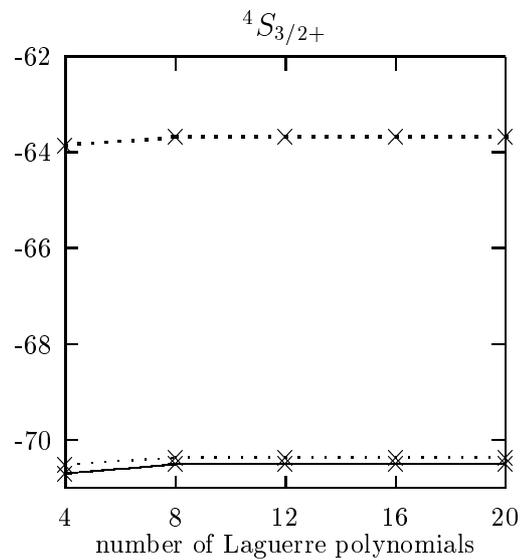
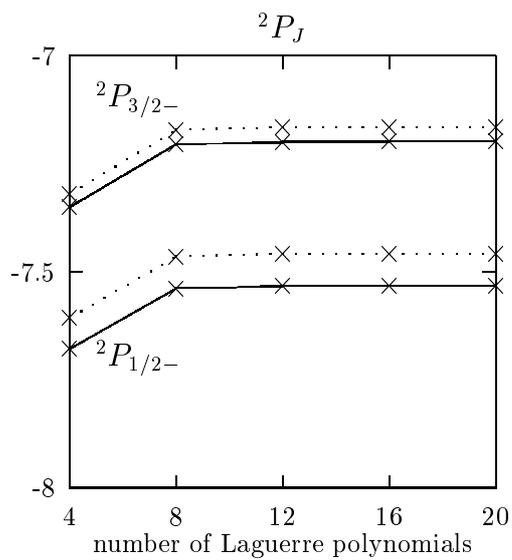
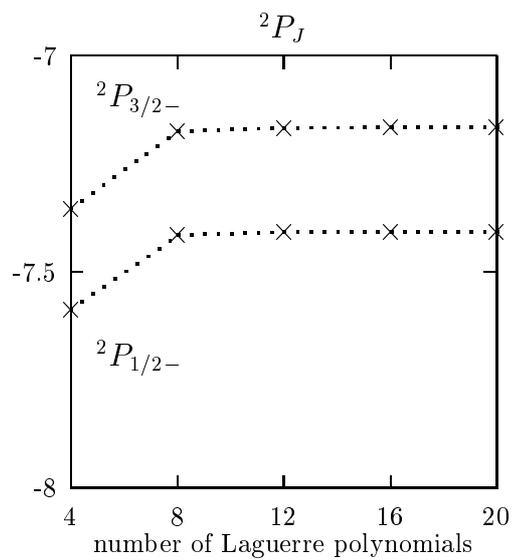
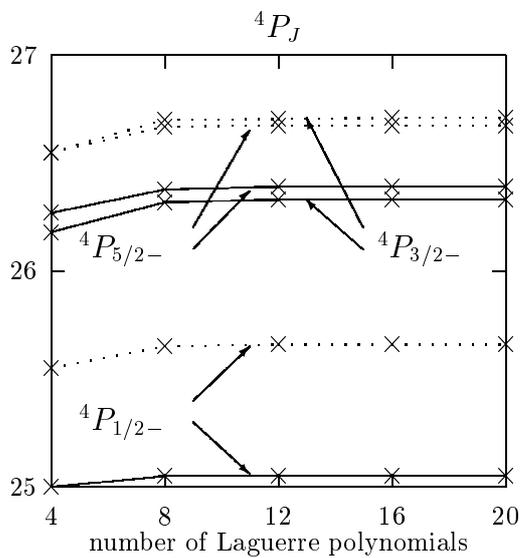
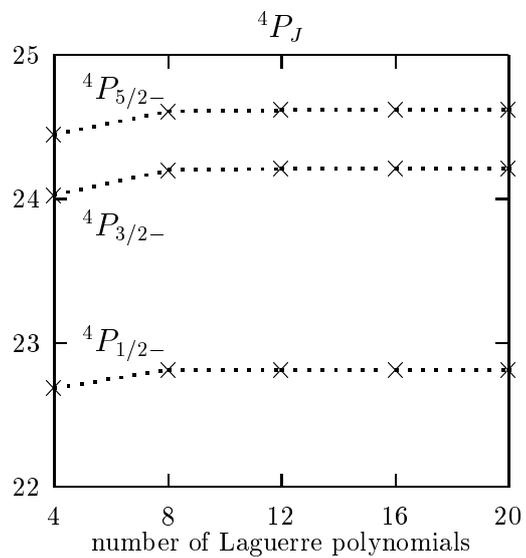
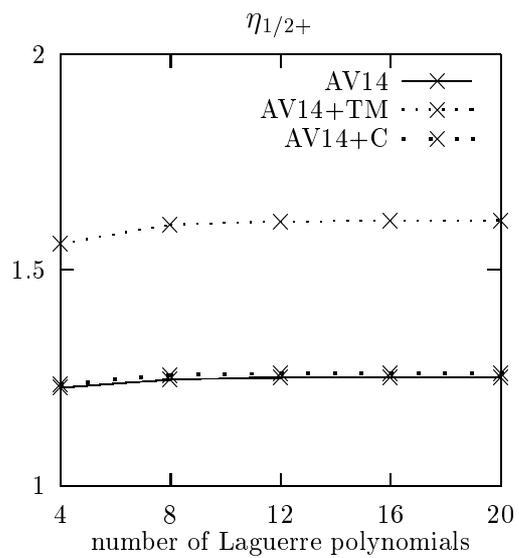
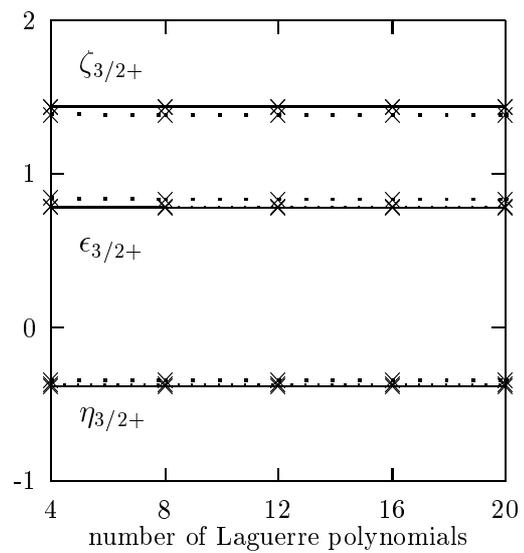
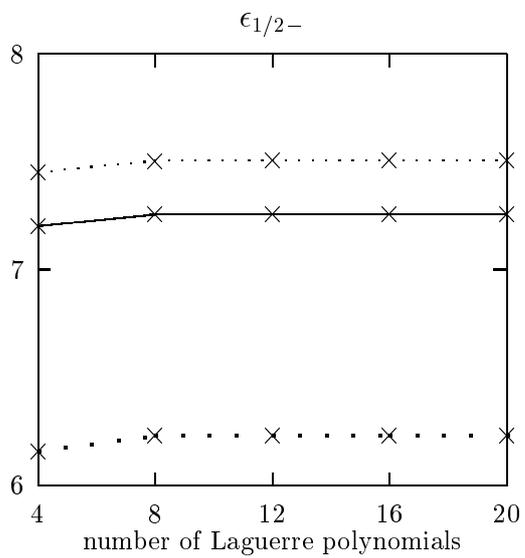
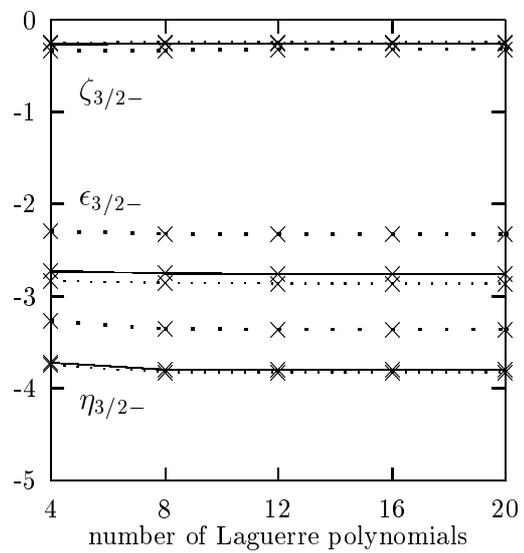
\end{document}